%% file: lum_bes3.tex
\documentclass[prd,showpacs,amsmath,amssymb,nofootinbib,twocolumn]{revtex4-1}
\RequirePackage{lineno}
\usepackage{graphicx}% Include figure files
\usepackage{dcolumn}% Align table columns on decimal point
\usepackage{bm}% bold math
\usepackage{subfigure}
\usepackage{epsfig}
\usepackage{epstopdf}
\usepackage{float}
\usepackage{times}
\usepackage{setspace}

\begin{document}
\normalsize
\parskip=5pt plus 1pt minus 1pt
%\linenumbers
\title{\bf  Precision measurement of the integrated luminosity of the 
data taken by BESIII at center of mass energies between 3.810 GeV and 
4.600 GeV}  

\input{authors_mar2015.tex}

\begin{abstract}
  From December 2011 to May 2014, about 5~$\rm fb^{-1}$ of data were
  taken with the BESIII detector at center-of-mass energies between
  3.810 GeV and 4.600 GeV to study the charmoniumlike states and higher
  excited charmonium states. The time-integrated luminosity of the
  collected data sample is measured to a precision of 1\% by analyzing
  events produced by the large-angle Bhabha scattering process.
\end{abstract}

\maketitle

\section{Introduction}

As a $\tau$-charm factory, the BESIII experiment has collected the
world's largest sample of $e^+ e^-$ collision data at
center-of-mass~(CM) energies between 3.810~GeV and~4.600 GeV.  In this
energy region, the charmoniumlike states and higher excited charmonium
states are produced copiously, which makes comprehensive studies
possible.

The charmoniumlike states discovered in recent years have drawn great
attention of both theorists and experimentalists for their exotic
properties, as reviewed \emph{e.g.} in Ref.~\cite{Brambilla:2010cs}. Being well
above the open charm threshold, the strong coupling of these states to
hidden charm processes makes their interpretation as conventional
charmonium states very difficult. On the other hand, the theory of the
strong interaction, Quantum Chromodynamics~(QCD), does not prohibit
the existence of exotic states beyond the quark model, \emph{e.g.} molecular
states, tetraquark states, hybrid states, \emph{etc}.  Either the
verification or the exclusion of the existence of such states will
help to evaluate the quark model and better understand QCD. Even
though some states have been identified as higher excited charmonium
states, such as the $\psi(4040)$, $\psi(4160)$, and $\psi(4415)$,
their large widths and the interference with each other make their
precise study complicated. In addition, the relationship between the
charmoniumlike states and higher excited charmonium states is still
not clear.  The precise knowledge of the time-integrated luminosity is
essential for quantitative analysis of these states.

In this paper, we present a measurement of the integrated luminosity 
based on the analysis of the Bhabha scattering process
\mbox{$e^+e^-\rightarrow (\gamma) e^+ e^-$}. A similar method has been
used in the luminosity measurement of $\psi(3770)$ data at
BESIII~\cite{3773}.  The process has a simple and clean signature and
a large production cross section, which allows for a small systematic
and a negligible statistical uncertainty. A cross check of the result
is performed by analyzing the di-gamma process \mbox{$e^+ e^-
\rightarrow\gamma\gamma$}.

\section{The detector}

BESIII is a general purpose detector which covers 93\% of the solid
angle and operates at the $e^+ e^-$ collider BEPCII. A detailed
description of the facilities is given in Ref.~\cite{BESIII}. The
detector consists of four main components: (a) A small-cell,
helium-based main drift chamber (MDC) with 43 layers provides an
average single-hit resolution of 135~$\mu$m, and a momentum resolution
of 0.5\% for charged tracks at 1~GeV/$c$ in a 1~T magnetic field. (b)
An electro-magnetic calorimeter~(EMC), consisting of 6240 CsI(Tl)
crystals in a cylindrical structure (barrel and two endcaps). The
energy resolution for 1.0~GeV photons is 2.5\%~(5\%) in the barrel~(endcaps), while the position resolution is 6~mm~(9~mm) in the barrel~(endcaps). (c) A time-of-fight system~(TOF), constructed of 5~cm thick
plastic scintillators, arranged in 88 detectors of 2.4~m length in two
layers in the barrel and 96 fan-shaped detectors in the endcaps. The
barrel (endcap) time resolution of 80~ps~(110~ps) provides 2$\sigma
K/\pi$ separation for momenta up to about 1.0~GeV/$c$. (d) A muon
counter~(MUC), consisting of nine layers of resistive plate chambers
in the barrel and eight layers for each endcap. It is incorporated in
the iron return yoke of the superconducting magnet. Its position
resolution is about 2~cm. A {\sc geant4}~\cite{geant41,geant42} based
detector simulation package has been developed to model the detector
response.  Due to the crossing angle of the beams at the
interaction point, the $e^+e^-$ CM system is slightly
boosted with respect to the laboratory frame.

\section{Data sample and Monte Carlo simulation}

Twenty-one data samples have been taken at CM energies between 3.810~GeV and 4.600~GeV. Six of the data sets exceed the others in
accumulated statistics by an order of magnitude. These samples were
taken on the peaks of charmoniumlike states, like the Y(4260),
Y(4360), and Y(4630), or higher excited charmonium states, like
$\psi(4040)$, and $\psi(4415)$, in order to study these resonances and
their decays in great detail. The data samples taken at the other CM
energies serve as scan points to study the behavior of the cross
section around these resonances. All individual data samples are
listed in Table~\ref{inf-Bhabha}. 

\begin{table*}[!htbp]
\centering \caption{\label{inf-Bhabha} Center-of-mass energy, 
  luminosity obtained from the nominal measurement~($L$),  cross check
  results~($L_{\rm ck}$), and  relative differences between the two
  results. The uncertainties are statistical only. Superscripts
  indicate separate samples acquired at the same CM energy.} 

\begin{tabular}{lrrc}
\hline
CM energy~(GeV)          &$L$ ($\rm pb^{-1}$) &$L_{\rm ck}$ ($\rm pb^{-1}$)  &Relative difference~(\%)\\
\hline
3.810                      &50.54$\pm$0.03       &50.11$\pm$0.08                 & $-0.85$$\pm$0.17\\
3.900                      &52.61$\pm$0.03       &52.57$\pm$0.08                 & $-0.08$$\pm$0.17\\
4.009                    &481.96$\pm$0.01     &480.54$\pm$0.23               & $-0.30$$\pm$0.05\\
4.090                      &52.63$\pm$0.03       &52.37$\pm$0.08                 & $-0.49$$\pm$0.17\\
4.190                      &43.09$\pm$0.03       &43.08$\pm$0.08                 & $-0.03$$\pm$0.20\\
4.210                      &54.55$\pm$0.03       &54.27$\pm$0.09                 & $-0.62$$\pm$0.18\\
4.220                      &54.13$\pm$0.03       &54.22$\pm$0.09                 & $+0.17$$\pm$0.18\\
4.230$^1$              &44.40$\pm$0.03       &44.64$\pm$0.08                 & $+0.54$$\pm$0.20\\
4.230$^2$              &1047.34$\pm$0.14   &1041.56$\pm$0.37             & $-0.56$$\pm$0.04\\
4.245                   &55.59$\pm$0.04       &55.52$\pm$0.09                 & $-0.13$$\pm$0.18\\
4.260$^1$              &523.74$\pm$0.10    &524.57$\pm$0.26               & $+0.16$$\pm$0.06\\
4.260$^2$             &301.93$\pm$0.08     &301.11$\pm$0.20               & $-0.28$$\pm$0.08\\
4.310                     &44.90$\pm$0.03       &45.29$\pm$0.08                 & $+0.87$$\pm$0.19\\
4.360                     &539.84$\pm$0.10     &541.38$\pm$0.28               & $+0.29$$\pm$0.06\\
4.390                     &55.18$\pm$0.04       &55.27$\pm$0.09                 & $+0.16$$\pm$0.18\\
4.420$^{1}$           &44.67$\pm$0.03       &44.77$\pm$0.08                 & $+0.22$$\pm$0.20\\
4.420$^{2}$           &1028.89$\pm$0.13    &1029.63$\pm$0.37            & $+0.07$$\pm$0.04\\
4.470                     &109.94$\pm$0.04      &109.51$\pm$0.13               & $-0.39$$\pm$0.13\\
4.530                     &109.98$\pm$0.04      &109.47$\pm$0.13               & $-0.46$$\pm$0.13\\
4.575                   &47.67$\pm$0.03        &47.57$\pm$0.08                & $-0.21$$\pm$0.18\\
4.600                     &566.93$\pm$0.11     &563.45$\pm$0.28                & $-0.62$$\pm$0.06\\
\hline
\end{tabular}
\end{table*}

At each energy point, one million Bhabha events were generated using the {\sc
  babayaga3.5}~\cite{babayaga} generator with the options presented in
Table~\ref{para}.  For the {\sc babayaga3.5} generator, the uncertainty in
calculating the cross section is 0.5\% , which meets the demand of the total
uncertainty of luminosity measurement. The kinematic distributions of the final state
particles from the {\sc babayaga3.5} generator are consistent with those from
data.  In the simulation, the scattering angles of the final state particles were
limited to a range from 20 degrees to 160 degrees, which slightly exceeds the
sensitive volume of the detector, in order to save on computing resources. An
energy threshold of 0.04~GeV was applied on the final state particles. The
acolinearity of the events has not been constrained. Finally, the generation
was taking into account the running of the electromagnetic coupling
constant and final state radiation~(FSR).

\begin{table}[!htbp]
\centering
\caption{\label{para} Options for the {\sc babayaga3.5} generator used to generate
  the simulated MC data samples.}
\begin{tabular}{lcc}
\hline
Parameters & Value \\
\hline
Ebeam                & 2.130 GeV or others   \\
MinThetaAngle        & 20$^{\circ}$         \\
MaxThetaAngle        & 160$^{\circ}$        \\
MinimumEnergy        & 0.04 GeV             \\
MaximumAcollinearity & 180$^{\circ}$        \\
RunningAlpha         &  1                   \\
FSR switch           &  1              \\
\hline
\end{tabular}
\end{table}

To study the background and optimize the event selection criteria, an
inclusive Monte Carlo~(MC) sample corresponding to a luminosity of
500~$\rm pb^{-1}$ at the CM energy of 4.260~GeV was generated, in which
the QED processes, the continuum production of hadrons, and the
initial state radiation~(ISR) to $J\slash\psi$ and $\psi(3686)$
resonance process were included.  
The {\sc babayaga3.5} generator was used to simulate the QED processes,
signal and background. Other processes, such as the decays of the
$J\slash\psi$, were generated with specialized models that have been
packaged and customized for the BESIII Offline Software
System~(BOSS)~(see \cite{prg} for an overview). 

\section{Event selection and result}

Signal candidates are required to have exactly two oppositely charged
tracks.  The tracks must originate from a cylindrical volume, centered
around the interaction point, which is defined by a radius of 1~cm
perpendicular to the beam axis and a length of $\pm10$~cm along the beam
axis.  In addition, the charged tracks are required to be within
$|\cos\theta|<0.8$, where $\theta$ is the polar angle, measured by the
MDC. Without applying further particle identification, the tracks are
assigned as electron and positron depending on their charge. The
deposited energies of electron and positron in EMC must be larger than
$\frac{\sqrt{s}}{4.26} \times 1.55$~(GeV) to remove the di-muon
background, where $\sqrt{s}$ is the CM energy in GeV; the momenta of
electron and positron are required to be larger than
$\frac{\sqrt{s}}{4.26} \times 2$~(GeV/c), to suppress background
events from lighter vector resonances produced in the ISR process,
such as $J\slash\psi$, $\psi(3686)$ and other resonances, decaying
into $e^{+}e^{-}$ pairs. For the data sample with a CM energy of 3.810
or 3.910 GeV, the effect of the remaining $\psi(3686)$ events is
studied by applying a 20\% larger momentum requirement, and is
found to be negligible. The requirements on the deposited energies and
momenta are not optimized in detail, as the number of the signal
events in such an analysis is large enough. All the variables
mentioned above are determined in the initial $e^{+}e^{-}$ CM
frame. The ratio of the number of remaining background events to
the number of signal events, estimated from the inclusive MC sample,
is found to be less than $2\times10^{-4}$, which is negligible. Thus
all the selected events are taken as Bhabha events. 

\begin{figure*}[!htbp]
  \centering
    \subfigure{
    \label{4260_1} %% label for first subfigure
    \includegraphics[width=0.31\textwidth]{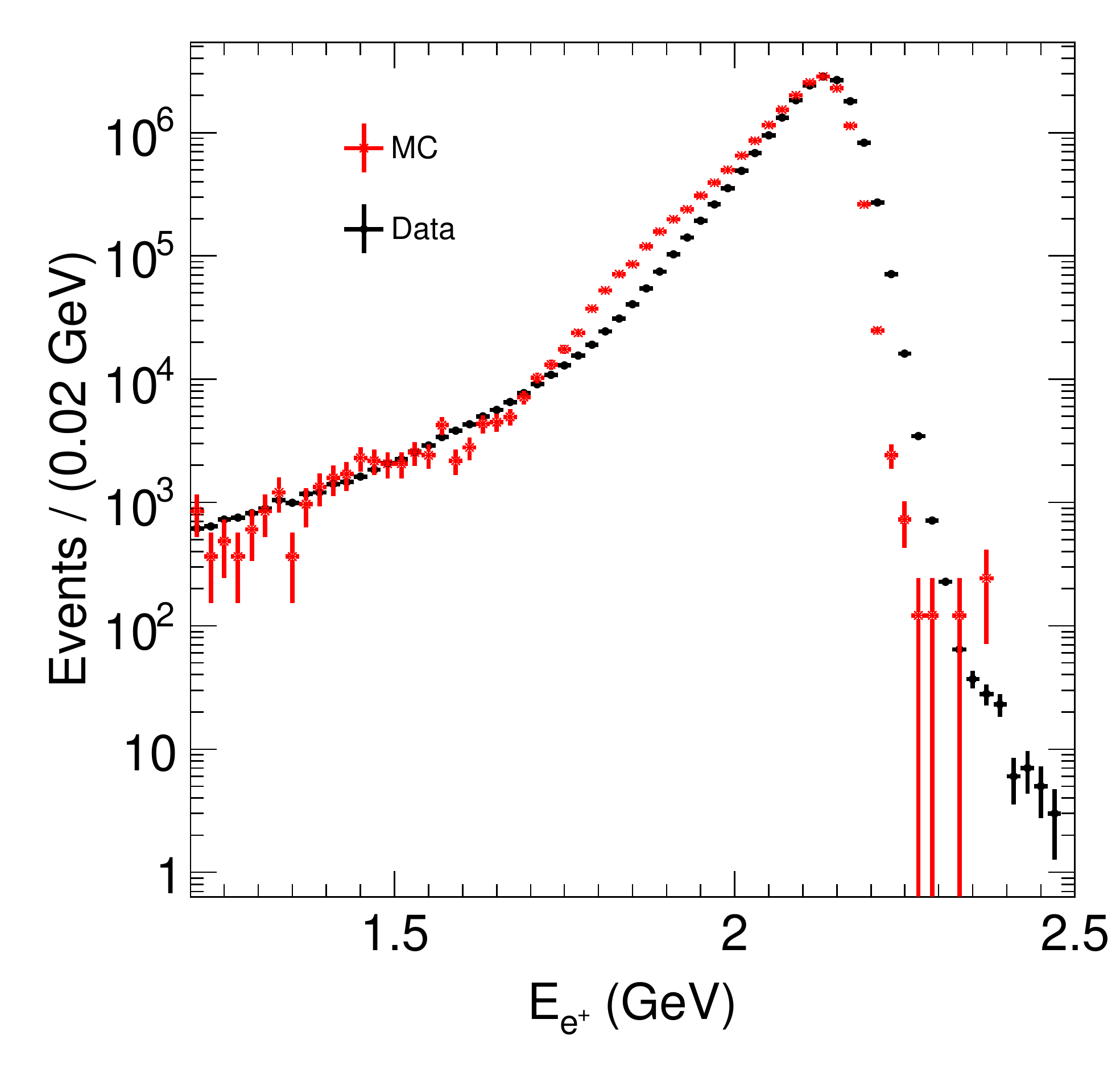}}
    \subfigure{
    \label{4260_3} %% label for first subfigure
    \includegraphics[width=0.31\textwidth]{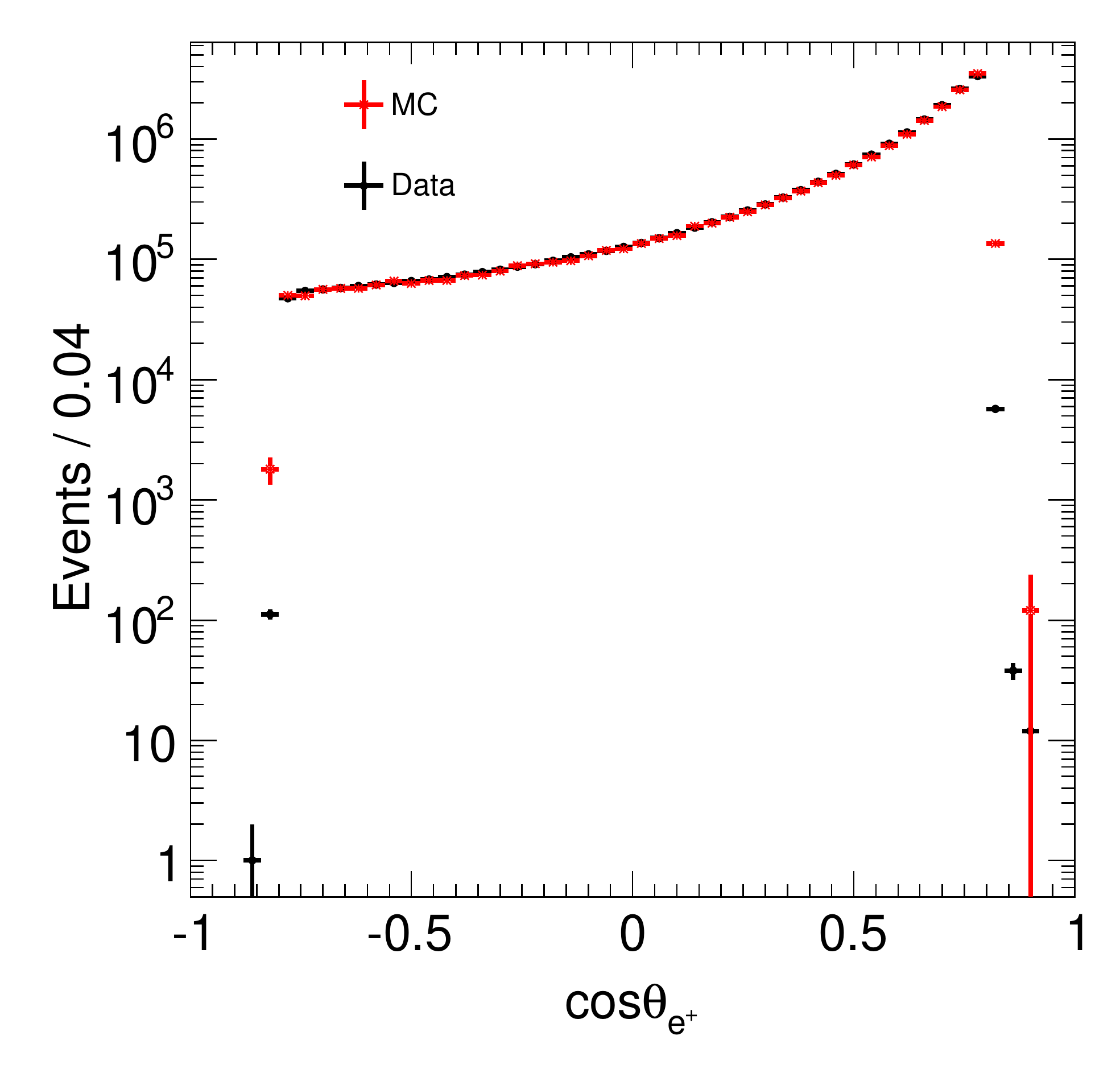}}
    \subfigure{
    \label{4260_5} %% label for first subfigure
    \includegraphics[width=0.31\textwidth]{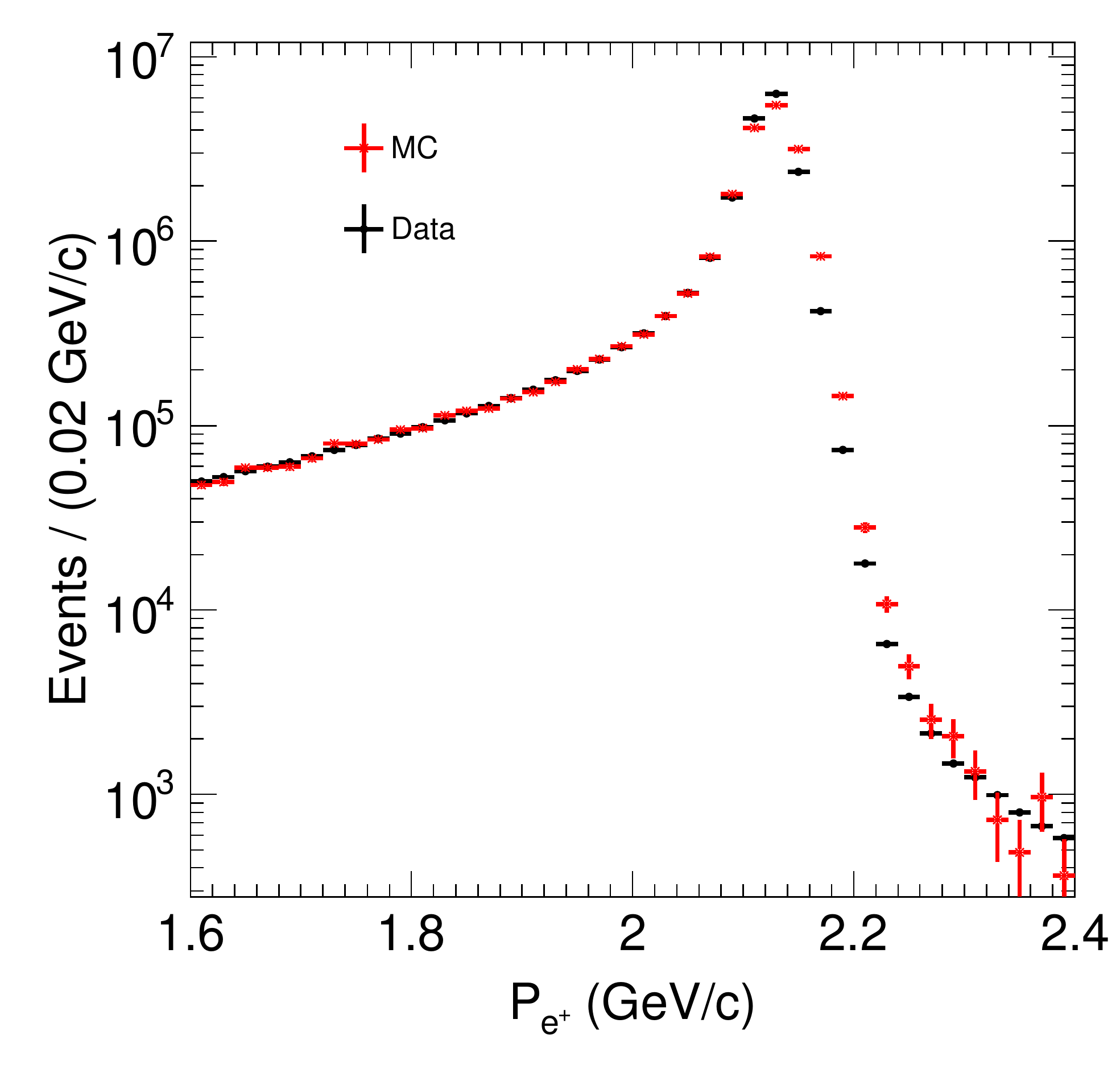}}
    \subfigure{
    \label{4260_2} %% label for first subfigure
    \includegraphics[width=0.31\textwidth]{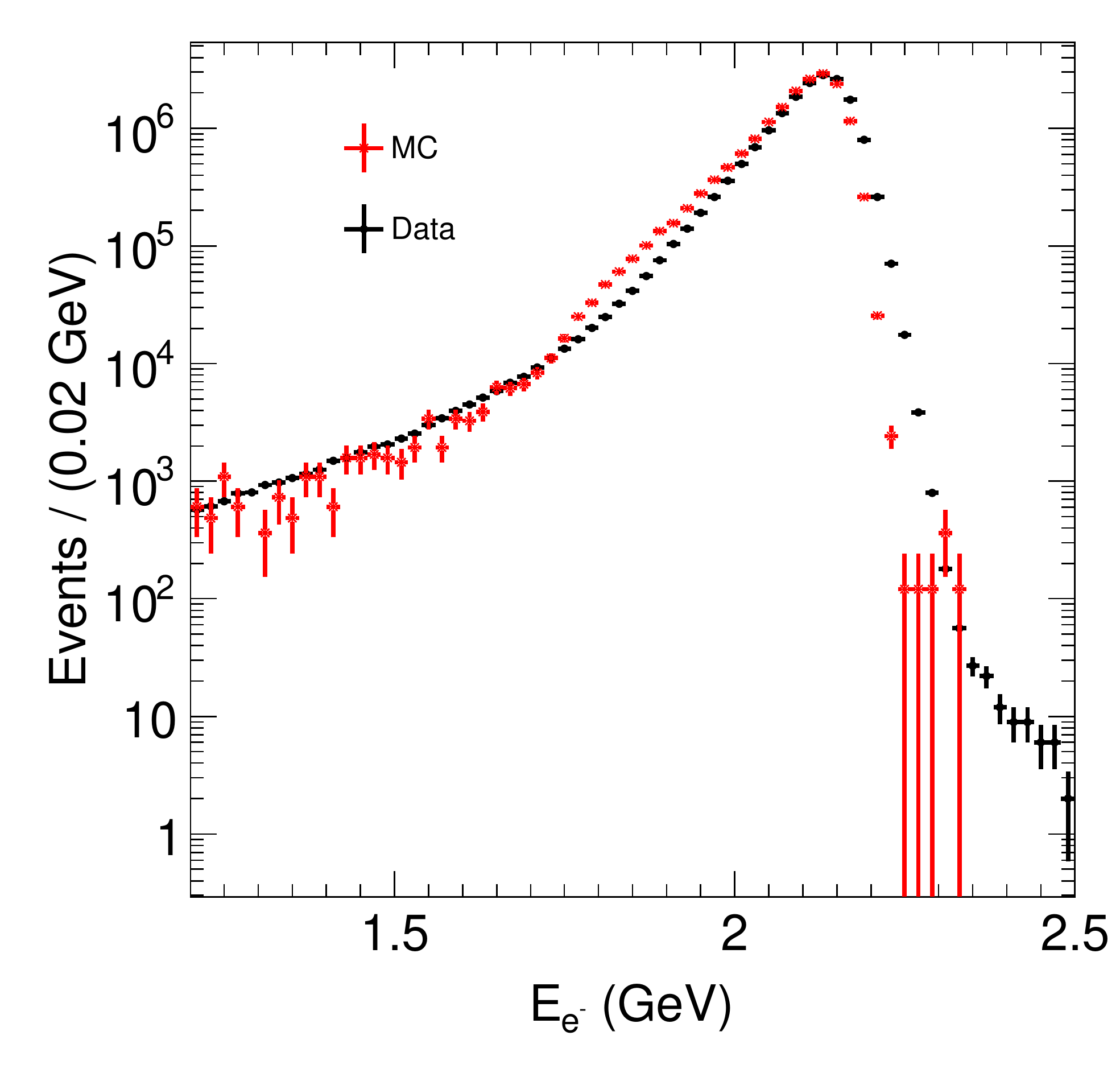}}
    \subfigure{
    \label{4260_4} %% label for first subfigure
    \includegraphics[width=0.31\textwidth]{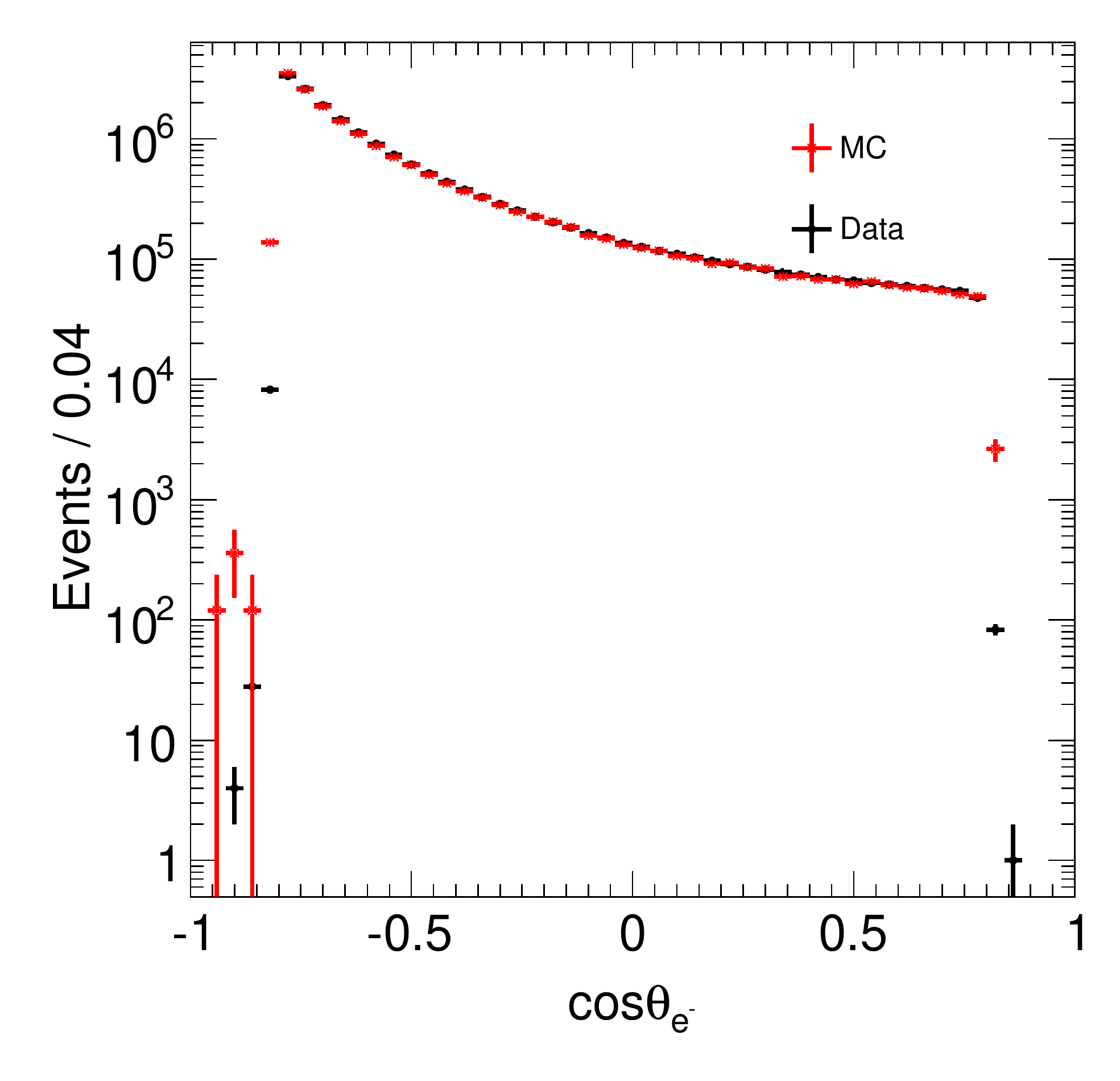}}
    \subfigure{
    \label{4260_6} %% label for first subfigure
    \includegraphics[width=0.31\textwidth]{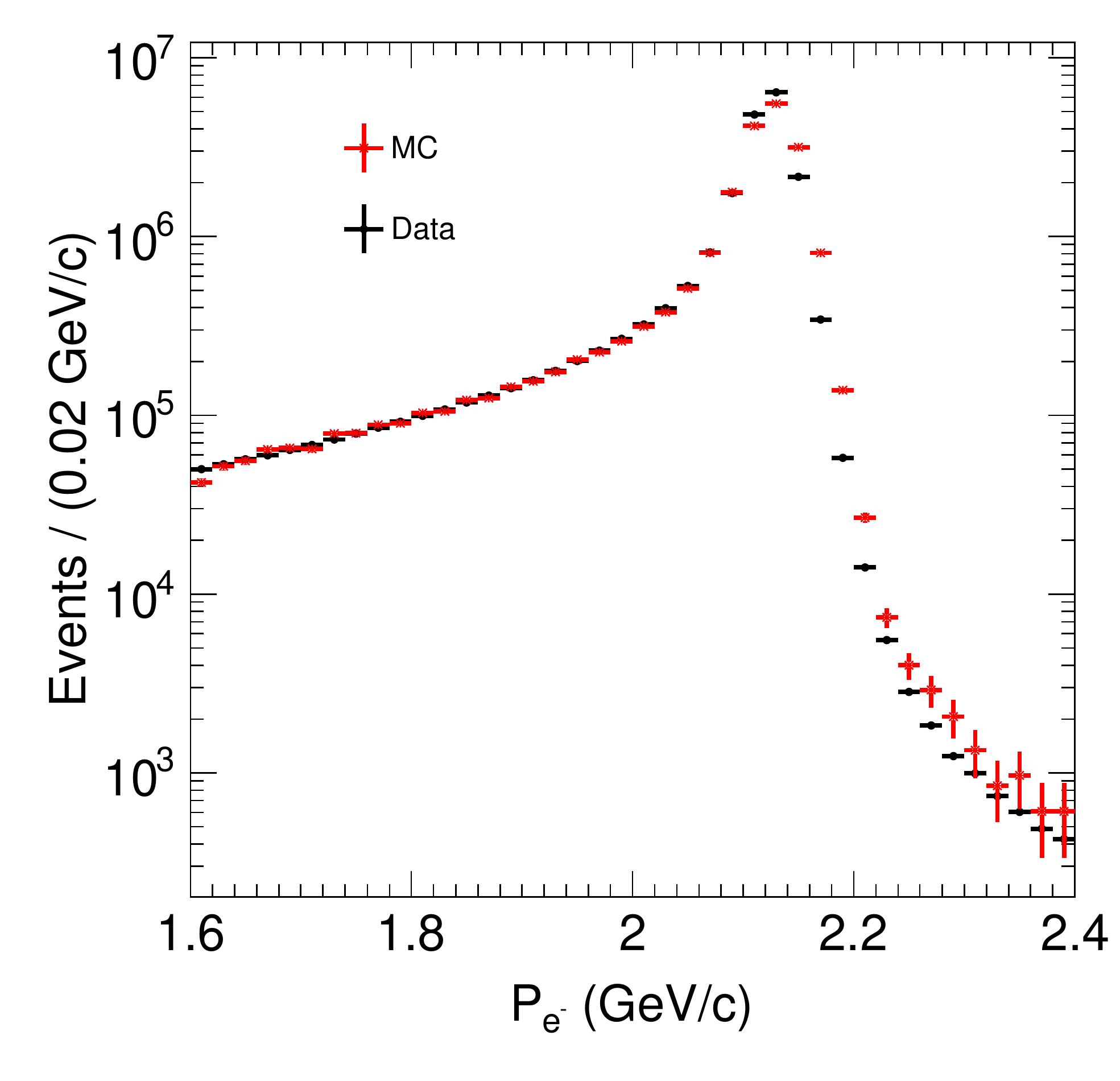}}
  \caption{Comparison between data and MC simulation at the CM
    energy of 4.260~GeV.  The top row is for positron and the bottom
    row for electron. From left to right, the plots show the
    distribution of deposited energy in EMC, the distribution of the
    cosine of the polar angle measured by the MDC, and the
    distribution of the track momentum from the MDC. Black dots with
    error bars illustrate data and the red ones are MC simulation.
    Note that the y-axis is in logarithmic scale and the MC is
    normalized to data by the number of events for each sub-plot.
    When drawing the distribution of one variable, the requirements on
    the other variables are applied.}
  \label{4260} %% label for entire figure
\end{figure*}

Figure~\ref{4260}
shows the comparisons between data and MC simulation for the kinematic
variables of the leptons by taking data at the CM energy of 4.260~GeV
as an example. Reasonable agreement is observed in the angular and
momentum distributions. The striking difference between data and
simulation found in the distributions of energies deposited by the
leptons in the EMC emerges from imperfections in the simulation of
the energy response of individual detector channels. At the CM
energies analyzed in this work, a single shower in the calorimeter can
be so energetic that the deposited energy per crystal exceeds the
dynamic range of the analog-to-digital converter~(ADC), causing
individual ADC channels to saturate.  
In the analysis presented in this report, conditions applied on the 
energy deposits are not affected.  Relevant deviations between data 
and MC are considered as contributions to the systematic uncertainties.

The integrated luminosity is calculated with
\begin{eqnarray}
\label{eq1}
L &=& \frac{N^{\rm obs}_{\rm Bhabha}}{\sigma_{\rm Bhabha}\times\epsilon},
\end{eqnarray}
where $N^{\rm obs}_{\rm Bhabha}$ is the number of observed Bhabha
events, $\sigma_{\rm Bhabha}$ is the cross section of the Bhabha
process, and $\epsilon$ is the efficiency determined by analyzing the
signal MC sample. The cross sections are calculated with the 
{\sc babayaga3.5} generator using the parameters listed in
Table~\ref{para} and decrease with increasing energies. The
efficiencies are almost independent of the CM energy, as intended by
the choice of relative conditions on lepton momenta and deposited
energies.  The luminosity results calculated with Equation~\ref{eq1}
are listed in Table~\ref{inf-Bhabha}.  The statistical accuracy of the
resulting integrated luminosity is better than 0.1\% at all energy
points.

\section{Systematic uncertainty}

The following sources of systematic uncertainties are considered: the
uncertainty of the tracking efficiency, the uncertainty related to the
requirements on the kinematic variables, the statistical uncertainty
of the MC sample, the uncertainty of the beam energy measurement, the
uncertainty of the trigger efficiency, and the systematic uncertainty
of the event generator.

To estimate the systematic uncertainty related to the tracking
efficiency, the Bhabha event sample is selected using information from
the EMC only, without using the tracking information in the MDC. The selection
criteria are: at least two clusters in the EMC for each candidate, and
the two most energetic clusters are assumed to originate from the $\rm
e^{+}e^{-}$ pair; the deposited energies of the two clusters are
required to be larger than $\frac{\sqrt{s}}{4.26}\times 1.8$~(GeV). At
CM energies above 4.420~GeV, the requirement is changed to
$\frac{\sqrt{s}}{4.26}\times 1.55$~(GeV). This adjustment allows to
avoid additional systematic uncertainties which would be introduced by the deviation
of data and simulation in the deposited energy in the EMC, as
discussed in Sec.IV.  The polar angle of each cluster is required to be
within $|\rm cos \theta^{EMC}|<0.8$, where $\theta^{\rm EMC}$ is the
polar angle measured by the EMC; to remove the background from the
di-photon process, $\Delta\phi$ is required to be in the range of
[$-40^\circ,-5^\circ$] or [$5^\circ,40^\circ$], where
$\Delta\phi=|\phi_{1}-\phi_{2}|-180^\circ$ and $\phi_{1,2}$ are
the azimuthal angles of the clusters in the EMC boosted to the CM
frame.  The efficiency that the selected Bhabha events
pass through the track requirements applied in the nominal analysis is
calculated for both data and MC sample, and the difference between
them is taken as the systematic uncertainty connected to the tracking
efficiency.

The systematic uncertainty in the requirement on the polar angle is
estimated by changing the requirement from \mbox{$|\rm
  cos\theta|<0.8$} to $|\rm cos\theta|<0.7$. The difference between
the resulting luminosity and nominal one is taken as the associated
systematic uncertainty. The systematic uncertainty caused by the
requirement on the energy deposited in the EMC is estimated by
changing the requirement from $\frac{\sqrt{s}}{4.26} \times
1.55$~(GeV) to $\frac{\sqrt{s}}{4.26} \times 1.71$~(GeV). The
systematic uncertainty caused by the requirement on the momentum is
estimated by changing the requirement from $\frac{\sqrt{s}}{4.26}
\times 2$~(GeV/c) to $\frac{\sqrt{s}}{4.26} \times 2.06$~(GeV/c). The
ranges are picked as these cause the largest deviations from the
nominal luminosity result near the requirements applied.

The statistical uncertainty of the efficiency determined from MC
simulations is 0.25\%. The CM energy is determined using
\mbox{$e^+e^-\rightarrow (\gamma) \mu^+ \mu^-$} events. The invariant mass of
the di-muon system is calculated taking into account ISR and FSR
effects~\cite{cms}. The difference between the CM energy listed in
Table~\ref{inf-Bhabha} and the one measured with di-muon process is
about 2~MeV, and the corresponding systematic uncertainty is estimated
by changing the CM energy by 2~MeV in the MC simulation. The trigger
efficiency for the Bhabha process is 100\% with an uncertainty of less
than 0.1\%~\cite{tri}. The theoretical uncertainty of the cross
section calculated by the {\sc babayaga3.5} generator is given as
0.5\%~\cite{babayaga}.

The same systematic uncertainty estimation method is applied to all
the sub-samples. The largest relative uncertainty among them is taken
as the associated uncertainty for all the sub-samples. The systematic
uncertainties considered in this work are summarized in
Table~\ref{sys}. By assuming the sources of the systematic
uncertainties to be uncorrelated, the total uncertainty is calculated
as 0.97\% by adding the contributions in quadrature.

\begin{table}[!htbp]
\centering
\caption{ \label{sys} Summary of the systematic uncertainties.}
\begin{tabular}{lc}
\hline
Source & Relative uncertainty (\%) \\
\hline
Tracking efficiency&0.39\\
Energy requirement &0.09\\
Momentum requirement &0.43\\
Polar angle requirement &0.38\\
MC statistics      &0.25\\
Beam energy        &0.42\\
Trigger efficiency &0.10\\
Generator          &0.50\\
\hline
Total              &0.97\\
\hline
\end{tabular}
\end{table}

\section{Cross check}

To verify the result, a cross check with di-gamma events is
performed. The event selection criteria are the same as those used in
estimating the systematic uncertainty caused by the tracking efficiency,
except for the requirement on $\Delta\phi$.  In order to reduce the Bhabha
background, the $\Delta\phi$ is required to be in the range of
[$-0.8^\circ,0.8^\circ$], since photons are not deflected in the
magnetic field.

The luminosity results of this cross check~($L_{\rm ck}$) are shown in
Table~\ref{inf-Bhabha}, together with the relative differences to the
nominal ones.  Both results are well consistent for all individual
measurements, indicating the robustness of the result.

\section{Summary}

The integrated luminosity of the data samples taken at BESIII for studying the
charmoniumlike states and higher excited charmonium states is
measured to a precision of 1\% with Bhabha events. The total
uncertainty is dominated by the systematic uncertainty. A cross check
with di-gamma events is performed and the results are consistent with
each other.  The result presented here is essential for future
measurements of cross sections with these data, and it has already
been used in the discovery of charged charmoniumlike
states~\cite{3900,4020,3885,4025}.

\begin{acknowledgements}
The BESIII collaboration would like to thank the staff of BEPCII and
the IHEP computing center for their strong support. This work is
supported in part by National Key Basic Research Program of China
under Contract No.~2015CB856700; National Natural Science Foundation
of China (NSFC) under Contracts Nos.~11125525, 11235011, 11322544,
11335008, 11425524; the Chinese Academy of Sciences (CAS) Large-Scale
Scientific Facility Program; Joint Large-Scale Scientific Facility
Funds of the NSFC and CAS under Contracts Nos.~11179007, U1232201,
U1332201; CAS under Contracts Nos.~KJCX2-YW-N29, KJCX2-YW-N45; 100
Talents Program of CAS; INPAC and Shanghai Key Laboratory for Particle
Physics and Cosmology; German Research Foundation DFG under Contract
No.\ Collaborative Research Center CRC-1044; Istituto Nazionale di
Fisica Nucleare, Italy; Ministry of Development of Turkey under
Contract No.~DPT2006K-120470; Russian Foundation for Basic Research
under Contract No.~14-07-91152; U.S.\ Department of Energy under
Contracts Nos.\ DE-FG02-04ER41291, DE-FG02-05ER41374,
DE-FG02-94ER40823, DESC0010118; U.S.\ National Science Foundation;
University of Groningen (RuG) and the Helmholtzzentrum fuer
Schwerionenforschung GmbH (GSI), Darmstadt; WCU Program of National
Research Foundation of Korea under Contract No. R32-2008-000-10155-0.
\end{acknowledgements}

\clearpage
\end{document}

%% file: authors_mar2015.tex
\author{
\begin{small}
\begin{center}
M.~Ablikim$^{1}$, M.~N.~Achasov$^{9,a}$, X.~C.~Ai$^{1}$, O.~Albayrak$^{5}$, M.~Albrecht$^{4}$, D.~J.~Ambrose$^{44}$, A.~Amoroso$^{48A,48C}$, F.~F.~An$^{1}$, Q.~An$^{45}$, J.~Z.~Bai$^{1}$
% ...
\\
\vspace{0.2cm}
(BESIII Collaboration)\\
\vspace{0.2cm} {\it
$^{1}$ Institute of High Energy Physics, Beijing 100049, People's Republic of China\\
% ...
$^{4}$ Bochum Ruhr-University, D-44780 Bochum, Germany\\
$^{5}$ Carnegie Mellon University, Pittsburgh, Pennsylvania 15213, USA\\
% ...
$^{9}$ G.I. Budker Institute of Nuclear Physics SB RAS (BINP), Novosibirsk 630090, Russia\\
% ...
$^{44}$ University of Rochester, Rochester, New York 14627, USA\\
$^{45}$ University of Science and Technology of China, Hefei 230026, People's Republic of China\\
% ...
$^{48}$ (A)University of Turin, I-10125, Turin, Italy; (B)University of Eastern Piedmont, I-15121, Alessandria, Italy; (C)INFN, I-10125, Turin, Italy\\
% ...
$^{a}$ Also at the Novosibirsk State University, Novosibirsk, 630090, Russia\\
% ...
}
\end{center}
\vspace{0.4cm}
\end{small}
}
\affiliation{}

%% file: lum_bes3.bbl
\begin{thebibliography}{90}
\bibitem{Brambilla:2010cs}
  N.~Brambilla, S.~Eidelman, B.~K.~Heltsley, R.~Vogt, G.~T.~Bodwin, E.~Eichten, A.~D.~Frawley and A.~B.~Meyer {\it et al.},
  Eur.\ Phys.\ J.\ C {\bf 71} 1534 (2011).
\bibitem{3773}
  M.~Ablikim {\it et al.}  [BESIII Collaboration],
  Chin.\ Phys.\ C {\bf 37}, 032007 (2013).
\bibitem{BESIII}
  M.~Ablikim {\it et al.}  [BESIII Collaboration],
  Nucl.\ Instrum.\ Meth.\ A {\bf 614}, 345 (2010).
\bibitem{geant41}
  S.~Agostinelli {\it et al.}  [GEANT4 Collaboration],
  Nucl.\ Instrum.\ Meth.\ A {\bf 506}, 250 (2003).
\bibitem{geant42}
  J.~Allison, K.~Amako, J.~Apostolakis, H.~Araujo, P.~A.~Dubois, M.~Asai, G.~Barrand and R.~Capra {\it et al.},
  IEEE Trans.\ Nucl.\ Sci.\  {\bf 53}, 270 (2006).
\bibitem{babayaga}
  G.~Balossini, C.~M.~Carloni Calame, G.~Montagna, O.~Nicrosini and F.~Piccinini,
  Nucl.\ Phys.\ B {\bf 758}, 227 (2006).
\bibitem{prg}
  R.~G.~Ping,
  Chin.\ Phys.\ C {\bf 32}, 599 (2008).
\bibitem{cms}
  M.~Ablikim {\it et al.}  [BESIII Collaboration], \emph{Measurement
    of the center-of-mass energy of the data taken by BESIII at center of mass energies between 3.81 GeV and 
4.60 GeV }, Paper in preparation.
\bibitem{tri}
  N.~Berger, K.~Zhu, Z.~-A.~Liu, D.~-P.~Jin, H.~Xu, W.~-X.~Gong, K.~Wang and G.~-F.~Cao,
  Chin.\ Phys.\ C {\bf 34}, 1779 (2010).
\bibitem{3900}
  M.~Ablikim {\it et al.}  [BESIII Collaboration],
  Phys.\ Rev.\ Lett.\  {\bf 110}, 252001 (2013).
\bibitem{4020}
  M.~Ablikim {\it et al.}  [BESIII Collaboration],
  Phys.\ Rev.\ Lett.\  {\bf 111}, 242001 (2013).
\bibitem{3885}
  M.~Ablikim {\it et al.}  [BESIII Collaboration],
  Phys.\ Rev.\ Lett.\  {\bf 112}, 022001 (2014).
\bibitem{4025}
  M.~Ablikim {\it et al.}  [BESIII Collaboration],
  Phys.\ Rev.\ Lett.\  {\bf 112}, 132001 (2014).
\end{thebibliography}
